\documentclass[aps,prd,nofootinbib,twocolumn,showpacs,superscriptaddress,
preprintnumbers,floatfix,notitlepage
]{revtex4-1}

\usepackage{amsmath}    
\usepackage{graphicx}   
\usepackage{verbatim}   
\usepackage{color}      
\usepackage{subfigure}  
\usepackage{bm}         
\usepackage{slashed}    
\usepackage{array}      
\usepackage{wasysym}    

\newcommand{\Feyn}[1]{#1\kern-0.45em/}

\newcommand{\Tr}[1]{\textnormal{Tr}\left[#1\right]}

\newcommand{\ket}[1]{\ensuremath{\left|{#1}\right\rangle}}

\newcommand{\strucFpart}[2]{F^{{#1}}_{{#2}}}
\newcommand{\strucF}[2]{{#1}\strucFpart{#1}{#2}}

\newcommand{\xbj}{x_B}

\usepackage{natbib}
\begin{document}

\title{Production of two hadrons in semi-inclusive Deep Inelatic Scattering
}

\author{S. Gliske}
\email{sgliske@umich.edu}
\affiliation{University of Michigan, Ann Arbor, Michigan 48109, USA}

\author{A. Bacchetta}
\email{alessandro.bacchetta@unipv.it}
\affiliation{Dipartimento di Fisica, Universit\`a di Pavia, via Bassi 6, I-27100 Pavia, Italy}
\affiliation{INFN, Sezione di Pavia, via Bassi 6, I-27100 Pavia, Italy}

\author{M. Radici}
\email{marco.radici@pv.infn.it}
\affiliation{INFN, Sezione di Pavia, via Bassi 6, I-27100 Pavia, Italy}

\date{\today}

\begin{abstract} 

We present the general expression, in terms of structure functions, of the cross section 
for the production of two hadrons in semi-inclusive deep-inelastic scattering. We analyze this process 
including full transverse-momentum dependence up to subleading twist and check, 
where possible, the consistency with existing literature.

\end{abstract} 

\maketitle

\section{Introduction}
\label{sec:intro}
Deep-inelastic lepton-nucleon scattering (DIS) is one of the key experimental 
tools to study the structure of nucleons, including their spin. Particularly
insightful can be semi-inclusive DIS (SIDIS), where one or more final-state
hadrons are detected. In this article, we take into consideration two-particle-inclusive 
DIS, i.e., DIS with two detected hadrons in the final state. And, in particular, those 
hadron pairs with low invariant mass.

One-particle-inclusive DIS has been studied in depth, including transverse-momentum 
dependence (see, e.g., \cite{Bacchetta:2006tn}). The analysis usually consists of two parts: 
1) a study of the general form of the cross section in terms of structure functions, which 
applies to any lepton-hadron scattering process with at least one hadron in the final state 
and relies on the single-photon-exchange approximation,  2) a study of the specific form 
of the structure functions in a parton-model framework. The second step has been carried 
out including corrections of order $M/Q$, where $M$ is the mass of the nucleon and $Q$ 
is the modulus of the transferred four-momentum. We shall call this level of approximation 
``subleading twist''. The underpinnings of the second step are factorization
theorems~\cite{Collins:1981uk,Ji:2002aa,Ji:2004wu,Collins:2011zzd,Aybat:2011zv,GarciaEchevarria:2011rb,Echevarria:2012pw,Collins:2012uy}, which however have been established only at the leading-twist level.
  
One can consider both transverse-momentum dependent (TMD) SIDIS, or collinear SIDIS, 
where all transverse momenta are integrated over. 
We present here both situations for the two-particle-inclusive case.

The production of two hadrons in SIDIS has been studied in several papers. 
The first comprehensive study has been presented in Ref.~\cite{Bianconi:1999cd} 
up to leading twist, where the relevant dihadron fragmentation functions have been
defined. Ref.~\cite{Bacchetta:2002ux} introduced the method of partial-wave analysis, 
which is important for our present discussion and clarifies the connection between 
two-hadron production and spin-one production~\cite{Bacchetta:2000jk}. 
Ref.~\cite{Bacchetta:2003vn} extended the analysis up to subleading twist, but only 
integrated over transverse momentum. Recent work has considered the problem of 
two-hadron production where one hadron is in the current region and one in the target
region~\cite{Anselmino:2011ss,Anselmino:2011bb,Anselmino:2012zz}.  The analysis has been
carried out at leading twist and is somewhat complementary to our present work.


This paper presents a slight modification to the definition of the fragmentation functions 
compared to, e.g., Ref.~\cite{Bacchetta:2003vn}. This not only may help in the interpretation 
and presentation of cross section moments, but it also has the practical advantage that 
the two-hadron SIDIS cross sections, at any twist, can be derived from single-hadron SIDIS.
Using this method, in this paper we present for the first time the expression of the TMD two-hadron 
SIDIS cross section at subleading twist including transverse-momentum dependence. We
cross-checked our result with existing literature for specific cases.

The paper is organized as follows. In Sec.~\ref{sec:cross_section}, we describe our notation, the kinematics, 
and we list the general expression for cross section in terms of structure functions. 
In Sec.~\ref{sec:newDiFF}, we describe our new more general definition of the transverse-momentum 
dependent two-hadron fragmentation functions, including their new partial-wave expansion. 
In Sec.~\ref{sec:struct_functs}, the structure functions are mapped onto specific convolutions involving 
the TMD distributions and two-hadron fragmentation functions. In Sec.~\ref{sec:specific_final_states}, 
we compare with known results for specific cases in order to clarify our nomenclature. 
Finally, some conclusions are drawn in Sec.~\ref{sec:conclusion}.

\section{Cross section in terms of structure functions}
\label{sec:cross_section}
\subsection{Definitions}
\label{sec:IIA}

We consider the process
\begin{equation}
  \label{eq:2hsidis}
\ell(l) + N(P) \to \ell(l') + h_1(P_1)+h_2(P_2) + X \; ,
\end{equation}
where $\ell$ denotes the beam lepton, $N$ the nucleon target, and $h$ the produced hadron, 
and where four-momenta are given in parentheses. We work in the one-photon exchange 
approximation and neglect the lepton mass.  We denote by $M$ the mass of the nucleon { and 
by $S$ its polarization}. The final hadrons have masses $M_1$, $M_2$ and momenta $P_1$, 
$P_2$. We introduce the pair total momentum $P_h = P_1 + P_2$ and relative momentum 
$R = (P_1-P_2)/2$. The invariant mass of the pair is $P_h^2 = M_h^2$.

As usual we define $q = l - l'$, where $Q^2 = -q^2$ is the hard scale of the process. 
We introduce the variables
\begin{align}
  \label{eq:xyz}
\xbj &= \frac{Q^2}{2\,P\cdot q},
&
y &= \frac{P \cdot q}{P \cdot l},
&
z_h &= \frac{P \cdot P_h}{P\cdot q},
&
\gamma &= \frac{2 M \xbj}{Q} \; .
\end{align}
The longitudinal polarization factor for the beam will be denoted $\lambda_e$ and
$\alpha$ is the fine structure constant. 

Of particular relevance for our discussions are the angles involved in the process. 
Two different sets of transverse projections are usually taken into
consideration. In fact, we can define two different transverse planes: the first is 
perpendicular to $(P, q)$, and the projection of a generic 4-vector $V$ onto it will be 
denoted by $V_\perp$; the second one is perpendicular to $(P, P_h)$ and the projection 
is indicated by $V_T$. The corresponding projection operators, up to terms of 
order $M^4/Q^4$, turn out to be\footnote{We use the convention $\epsilon^{0123} = 1$.}
\begin{align}
g_\perp^{\mu\nu} &= g_{}^{\mu\nu} 
  - \frac{q^\mu P^\nu + P^\mu q^\nu}{P\cdot q\, (1+\gamma^2)}
  + \frac{\gamma^2}{1+\gamma^2} \left(
      \frac{q^\mu q^\nu}{Q^2} - \frac{P^\mu P^\nu}{M^2} \right)  \; ,
\\
\epsilon_\perp^{\mu\nu} &= \epsilon_{}^{\mu\nu\rho\sigma}
   \frac{P_\rho\, q_\sigma}{P\cdot q\, \sqrt{1+\gamma^2}}
\end{align} 
and
\begin{align} 
\begin{split} 
g_T^{\mu \nu}&=  g_{}^{\mu\nu}
    - \frac{2 \xbj}{Q^2 z_h} \Big(P^{\mu} P_h^{\nu} + P_h^{\mu} P^{\nu}\Big)
\\
& \quad + \frac{M_h^2 \gamma^2 }{Q^2 z_h^2} 
     \bigg(\frac{P^{\mu} P^{\nu}}{M^2}+\frac{P_h^{\mu} P_h^{\nu}}{M_h^2} \bigg) \; ,
\end{split} 
\\
\epsilon_T^{\mu \nu} &= \epsilon^{\mu \nu \rho \sigma}
               \frac{P_{\rho} P_{h\sigma}}{P \cdot P_h} \; .
\end{align} 

We define the azimuthal angles~\cite{Bacchetta:2004jz,Bacchetta:2006tn}
\begin{align}
  \label{eq:phi-h-def}
\cos\phi_h &= - \frac{l_\mu P_{h \nu}\, g_\perp^{\mu\nu}}{%
  \sqrt{ l_\perp^2\, P_{h \perp}^2}} \,,
&
\sin\phi_h &= - \frac{l_\mu P_{h \nu}\, \epsilon_\perp^{\mu\nu}}{%
  \sqrt{ l_\perp^2\, P_{h \perp}^2}} \,,
\end{align}
where $l_\perp^\mu = g_\perp^{\mu\nu} l^{}_\nu$ and $P_{h \perp}^\mu = g_\perp^{\mu\nu} P^{}_{h \nu}$.
The azimuthal angle of the spin vector, $\phi_S$, is defined in analogy to $\phi_h$, 
with $P_h$ replaced by $S$.

For dihadron fragmentation functions, we need to introduce one more azimuthal angle.
We first introduce the vector $R_T$, i.e., the component of $R$ perpendicular to $P$ and $P_h$. 
Defining the invariant
\begin{equation} 
\zeta_h= \frac{2 R \cdot P}{P_h \cdot P}\; , 
\end{equation} 
neglecting terms of order $M^4/Q^4$ we can write
\begin{equation}
\begin{split}
R_T^{\mu} &= g_T^{\mu \nu} R_{\nu} \\ 
&= R^{\mu} - \frac{\zeta_h}{2} P_h^{\mu} +\xbj \frac{\zeta_h M_h^2 - (M_1^2-M_2^2)}{Q^2 z_h} P^{\mu} \; .
\end{split} 
\label{eq:RT}
\end{equation} 
However, the cross section will depend on the azimuthal angle of $R_T$ measured in the plane 
perpendicular to $(P, q)$. Therefore, we need to use Eq.~\eqref{eq:phi-h-def} replacing $P_h$ 
with $R_T$. We will denote the azimuthal angle of $R_T$ in this frame by $\phi_{R_\perp}$. 
This choice is similar to what has been done in Ref.~\cite{Airapetian:2008sk}, but here it has 
been realized in a covariant way. In Appendix~\ref{sec:appendixA}, we compare our definition with 
other non-covariant ones available in the literature, pointing out the potential differences depending 
on the choice of the reference frame. 

It is anyway convenient to give the expression of the involved angles in specific frames of reference. 
The azimuthal angles are usually written in the target rest frame (or in any frame reached from 
the target rest frame by a boost along $\bm q$)
\begin{align} 
\phi_{h}&= 
\frac{ \left(\bm q\times\bm l\right)\cdot \bm P_h }{| \left(\bm q\times\bm l\right)\cdot \bm P_h |}  
\arccos \frac{ \left(\bm q \times \bm l\right) \cdot \left(\bm q \times \bm P_h \right)}
                                          {\left| \bm q \times \bm l\right| \left| \bm q \times \bm P_h \right| } \; ,
\\
\phi_{R_\perp}&= 
\frac{ \left(\bm q\times\bm l\right)\cdot \bm R_T }{| \left(\bm q\times\bm l\right)\cdot \bm R_T |}  
\arccos \frac{ \left(\bm q \times \bm l\right) \cdot \left(\bm q \times \bm R_T \right)}
                                          {\left| \bm q \times \bm l\right| \left| \bm q \times \bm R_T \right| } \; .
\label{eq:phi_Rperp_def}
\end{align}  

In the center-of-mass (cm) frame of the two hadrons, the emission occurs back-to-back 
and the key variable is the polar angle $\vartheta$ between the directions of the emission and 
of $P_h$~\cite{Bacchetta:2002ux}. The variable $\zeta_h$ can be written in terms of the $\vartheta$ 
as follows
\begin{equation} 
\begin{split}
|\bm R| &= \frac{1}{2} \sqrt{M_h^2 - 2(M_1^2+M_2^2) + (M_1^2-M_2^2)^2/M_h^2} \; , \\[2mm]
\zeta_h &= \frac{1}{M_h} \Big(\sqrt{M_1^2 - |\bm R|^2} - \sqrt{M_2^2 - |\bm R|^2} - 2
|\bm R| \cos \vartheta \Big) \; .
\end{split}
\label{eq:theta}
\end{equation}

At this point, we remark also that the analysis of two-hadron production can be done in terms 
of the variables ($z_{h1}$, $z_{h2}$, $P_{h1 T}$, $P_{h2 T}$), of the two individual 
hadrons, instead of introducing the sum and difference of their momenta. This choice is more 
reasonable if one hadron is in the current region and one in the target
region~\cite{Anselmino:2011ss,Anselmino:2011bb,Anselmino:2012zz}.

\subsection{Complete dependence of the cross section}
\label{sec:IIB}

The cross section is split in parts denoted by $\sigma_{XY}$, based on the target ($X$) 
and beam ($Y$) polarization, $X$ and $Y$ taking values $U$ (unpolarized), 
$L$ (longitudinally polarized) and $T$ (transversely polarized). The structure functions 
will likewise have subscripts $XY$, with the same meaning.  In a few cases the structure 
functions have an additional subscript, indicating a longitudinal ($L$) or transverse ($T$) 
virtual photon polarization.

We introduce the depolarization factors~\cite{Bacchetta:2006tn}
\begin{align} 
\begin{split} 
\label{eq:Axy}
  A(x,y) &= \frac{y^2}{2(1-\epsilon)}
        = \frac{1-y+\frac{1}{2}y^2 + \frac{1}{4}\gamma^2y^2}{1+\gamma^2}
\\
       & \approx  \left(1-y+\frac{1}{2}y^2\right) \; , 
\end{split} 
\\
\begin{split} 
\label{eq:Bxy}
  B(x,y) &= \frac{y^2}{2(1-\epsilon)}\epsilon
        = \frac{1-y-\frac{1}{4}\gamma^2y^2}{1+\gamma^2} \\
       & \approx  \left(1-y\right) \; , 
\end{split} 
\\
\begin{split} 
\label{eq:Cxy}
  C(x,y) &= \frac{y^2}{2(1-\epsilon)}\sqrt{1-\epsilon^2}
        =  \frac{y\left(1-\frac{1}{2}y\right)}{\sqrt{1+\gamma^2}} \\
       & \approx  y\left(1-\frac{1}{2}y\right) \; , 
\end{split} 
\\
\begin{split} 
\label{eq:Vxy}
  V(x,y) &= \frac{y^2}{2 (1-\epsilon)}\sqrt{2\epsilon(1+\epsilon)} \\
       &= \frac{2-y}{1+\gamma^2}\sqrt{1-y-\frac{1}{4}\gamma^2y^2} \\
       & \approx  \left(2-y\right)\sqrt{1-y}\; , 
\end{split} 
\\
\begin{split} 
\label{eq:Wxy}
  W(x,y) &= \frac{y^2}{2 (1-\epsilon)}\sqrt{2\epsilon(1-\epsilon)} \\
        &= \frac{y}{\sqrt{1+\gamma^2}}\sqrt{1-y-\frac{1}{4}\gamma^2y^2} \\
       & \approx  y\sqrt{1-y} \; .
\end{split} 
\end{align} 
The approximations, which no longer depend on $x$, are valid up to corrections
of order  $M^2/Q^2$.

The cross section will be differential in the following variables
\begin{equation}
\frac{d\sigma}{d\xbj \, dy\, d\psi \,dz_h\, d\phi_h\, d P_{h \perp}^2  d\phi_{R_\perp}\, d M_{h}\,d \cos{\theta}} \; .
\end{equation} 
The angle $\psi$ is the azimuthal angle of $\ell'$ around the lepton beam axis 
with respect to an arbitrary fixed direction, which in case of a transversely polarized target 
we choose to be the direction of $S$. The corresponding relation between $\psi$
and $\phi_S$ is given in Ref.~\cite{Diehl:2005pc}; neglecting corrections of order 
$M^2/Q^2$, one has $d \psi \approx d \phi_S$. 

The dependence of the cross section on the polar angle $\cos\vartheta$ and on the 
azimuthal angles $\phi_h, \, \phi_{R_\perp},$ is transformed by expanding it on a basis of 
spherical harmonics. In particular, for the $\cos\vartheta$ dependence we adopt the basis 
of Legendre polynomials, the first few of which read
\begin{align} 
  P_{0,0} &= 1\; ,        
&   
  P_{2,0} &= \frac{1}{2} \left( 3 \cos^2\vartheta - 1 \right)\; ,
\\
  P_{1,0} &= \cos\vartheta \; ,
&   
  P_{2,1} &= \sin2\vartheta \; ,
\\
  P_{1,1} &= \sin\vartheta \; ,
&   
  P_{2,2} &= \sin^2\vartheta \dots
\label{eq:Plm}
\end{align} 
with $P_{\ell,-m} = P_{\ell,m}$.



For the one-particle-inclusive SIDIS case, the hadronic tensor is built by using 3 four-vectors, 
$q,  P,  P_h, $ and 1 pseudo four-vector, $S$. Since the target is a spin-$1/2$ particle, 
the hadronic tensor can be at most linear in $S$. By imposing the invariance under the 
usual transformations (parity, time-reversal, gauge), the hadronic tensor can be parametrized 
in terms of 18  structure functions~\cite{Mulders:1995dh,Bacchetta:2006tn}. In the two-particle-inclusive SIDIS, 
even in the simplest case when the target and the two final hadrons are unpolarized, 
the pseudo-vector $S$ is replaced by $R$ and the hadronic tensor does not necessarily 
need to be linear in $R$. Actually, the number of partial waves depending on $\phi_{R_\perp}$ is 
in principle not limited, and so the number of structure functions is also not limited. 

The structure of the cross section for unpolarized beam and unpolarized target, is similar to 
the one for the one-particle-inclusive SIDIS, because it is dictated by the helicity density matrix of 
the virtual photon: there are two diagonal elements related to its transverse ($T$) and longitudinal 
($L$) polarization, and there are two interference terms. Then, in this cross 
section four different parts can be identified, each one displaying an infinite number of 
structure functions: 
\begin{widetext}
\begin{equation} 
\begin{split} 
\label{eq:sigma_UU}
  d\sigma_{UU} &= \frac{\alpha^2}{4 \pi xyQ^2}\left(1+\frac{\gamma^2}{2x}\right) \\
  &\quad \times \sum_{\ell=0}^{\ell_{\max}} \ \Bigg\{ \ A(x,y) \sum_{m=0}^{\ell} \bigg[ P_{\ell,m} \cos(m(\phi_h-\phi_{R_\perp}))
     \left( \strucFpart{P_{\ell,m}\cos(m(\phi_h-\phi_{R_\perp}))}{UU,T} +
                    \epsilon \strucFpart{P_{\ell,m}\cos(m(\phi_h-\phi_{R_\perp}))}{UU,L} \right ) \bigg] \\
  &\quad \hspace*{0.2in} {} + 
    B(x,y) \sum_{m=-\ell}^{\ell} \strucF{P_{\ell,m}\cos((2-m)\phi_h + m\phi_{R_\perp})}{UU}\\
  &\quad \hspace*{0.2in} {} + 
    V(x,y) \sum_{m=-\ell}^{\ell} \strucF{P_{\ell,m}\cos((1-m)\phi_h + m\phi_{R_\perp})}{UU} \Bigg\} \; . 
\end{split}
\end{equation} 
\end{widetext}
As explained above, there is no upper limit to $\ell_{\max}$: there are infinitely many azimuthal 
modulations, in contrast to the 18 structure functions of single-hadron production. In this case, 
what limits the number of structure functions is conservation of angular momentum, 
as discussed for instance in Ref.~\cite{Diehl:2005pc}. Here, however, the presence of 
two sources of angular momentum (the total and relative angular momenta of the pair) 
allows for infinitely many combinations. The only constraint is that the sum of the coefficients 
of $\phi_h$ and $\phi_{R_\perp}$ should be limited to at most 3: this is the maximum mismatch 
of angular momentum projections in the $\gamma$ $P$ system.

For practical purposes, in certain situations it may be possible to restrict the value of $\ell_{\max}$. 
For instance, when considering two hadrons emitted through a vector meson resonance, $\ell_{\max}=2$.  
However, it is not possible in general to distinguish between resonant and non-resonant dihadron production, 
e.g., between non-resonant $\pi^+\pi^-$ and resonant $\rho^0$. The non-resonant dihadrons are 
not necessarily restricted to any finite $\ell$ value, although at small invariant mass higher $\ell$ 
should be suppressed.

The structure functions on the r.h.s.\ of Eq.~\eqref{eq:sigma_UU} depend on $\xbj$, $Q^2$, $z_h$,
$P_{h \perp}^2$, $M_h^2$. 

Along the same lines, the cross section for longitudinally polarized beam and unpolarized target reads
\begin{widetext}
\begin{equation} 
\begin{split} 
\label{eq:sigma_LU}
  d\sigma_{LU} &= \frac{\alpha^2}{4 \pi xyQ^2}\left(1+\frac{\gamma^2}{2x}\right) \lambda_e \\
  &\quad \times \sum_{\ell=0}^{\ell_{\max}} \ \Bigg\{  
     \ C(x,y) \sum_{m=1}^{\ell} \bigg[ P_{\ell,m} \sin(m(\phi_h-\phi_{R_\perp})) \, 2 \, 
     \left( \strucFpart{P_{\ell,m}\cos(m(\phi_h-\phi_{R_\perp}))}{LU,T} +
                    \epsilon \strucFpart{P_{\ell,m}\cos(m(\phi_h-\phi_{R_\perp}))}{LU,L} \right ) \bigg] \\
  &\quad \hspace*{0.2in} {} + 
    W(x,y) \sum_{m=-\ell}^{\ell} \strucF{P_{\ell,m}\sin((1-m)\phi_h + m\phi_{R_\perp})}{LU}\Bigg\} \; .
\end{split} 
\end{equation} 

The cross section for unpolarized beam and longitudinally polarized target reads
\begin{equation} 
\begin{split} 
\label{eq:sigma_UL}
  d\sigma_{UL} &= \frac{\alpha^2}{4 \pi xyQ^2}\left(1+\frac{\gamma^2}{2x}\right) S_L \\
  &\quad \times \ \Bigg\{
    A(x,y) \sum_{\ell=1}^{\ell_{\max}} \sum_{m=1}^{\ell} \strucF{P_{\ell,m}\sin(-m\phi_h + m\phi_{R_\perp})}{UL}\\
  &\quad \hspace*{0.2in} {} + 
    B(x,y) \sum_{\ell=0}^{\ell_{\max}} \sum_{m=-\ell}^{\ell} \strucF{P_{\ell,m}\sin((2-m)\phi_h + m\phi_{R_\perp})}{UL}\\
  &\quad \hspace*{0.2in} {} + 
    V(x,y) \sum_{\ell=0}^{\ell_{\max}}\sum_{m=-\ell}^{\ell} \strucF{P_{\ell,m}\sin((1-m)\phi_h + m\phi_{R_\perp})}{UL} \Bigg\} \; . 
\end{split}
\end{equation} 

The cross section for longitudinally polarized beam and longitudinally polarized target reads 
\begin{equation}
\begin{split}
\label{eq:sigma_LL}
  d\sigma_{LL} &= \frac{\alpha^2}{4 \pi xyQ^2}\left(1+\frac{\gamma^2}{2x}\right) \lambda_e S_L \\
  &\quad \times \sum_{\ell=0}^{\ell_{\max}} \ \Bigg\{
      \ C(x,y) \sum_{m=0}^{\ell} 2^{2-\delta_{m0}} \,  \strucF{P_{\ell,m} \cos(m(\phi_h-\phi_{R_\perp}))}{LL}\\
  &\quad \hspace*{0.2in} {} + 
    W(x,y) \sum_{m=-\ell}^{\ell} \strucF{P_{\ell,m}\cos((1-m)\phi_h + m\phi_{R_\perp})}{LL} \Bigg\} \; .
\end{split} 
\end{equation}

The cross section for unpolarized beam and transversely polarized target reads 
\begin{equation}
\begin{split} 
\label{eq:sigma_UT}
  d\sigma_{UT} &= \frac{\alpha^2}{4 \pi xyQ^2}\left(1+\frac{\gamma^2}{2x}\right) |\bm{S}_\perp| \\
&\quad \times \ \sum_{\ell=0}^{\ell_{\max}}\sum_{m=-\ell}^{\ell} \Bigg\{ \ A(x,y) 
    \bigg[ P_{\ell,m} \sin((m+1)\phi_h-m\phi_{R_\perp}-\phi_S))
\\  &\quad \hspace*{0.4in} {} \times
 \left( \strucFpart{P_{\ell,m}\sin((m+1)\phi_h-m\phi_{R_\perp}-\phi_S)}{UT,T} +
                    \epsilon \strucFpart{P_{\ell,m}\sin((m+1)\phi_h-m\phi_{R_\perp}-\phi_S)}{UT,L} \right ) \bigg] \\
  &\quad \hspace*{0.2in} {} + 
    B(x,y)  \bigg[
      \strucF{P_{\ell,m}\sin((1-m)\phi_h + m\phi_{R_\perp} + \phi_S)}{UT}\\
  &\quad \hspace*{0.4in} {} + \strucF{P_{\ell,m}\sin((3-m)\phi_h + m\phi_{R_\perp} - \phi_S)}{UT}\bigg] \\
  &\quad \hspace*{0.2in} {} + 
    V(x,y) \bigg[ \strucF{P_{\ell,m}\sin( - m\phi_h + m\phi_{R_\perp} + \phi_S)}{UT} \\
  &\quad \hspace*{0.4in} {} + \strucF{P_{\ell,m}\sin((2-m)\phi_h + m\phi_{R_\perp} - \phi_S)}{UT}\bigg]  \Bigg\} \; .
\end{split} 
\end{equation}

Lastly, the cross section for longitudinally polarized beam and transversely polarized target reads 
\begin{equation}
\begin{split} 
\label{eq:sigma_LT}
  d\sigma_{LT} &= \frac{\alpha^2}{4 \pi xyQ^2}\left(1+\frac{\gamma^2}{2x}\right) \lambda_e |\bm S_\perp|
\sum_{\ell=0}^{\ell_{\max}}\sum_{m=-\ell}^{\ell} \Bigg\{ \\ 
&\quad \hspace*{0.2in} 
       \ C(x,y) \, 2 \,  \strucF{P_{\ell,m} \cos((1-m)\phi_h+m\phi_{R_\perp}-\phi_S))}{LT}\\
  &\quad \hspace*{0.2in} {} + 
         W(x,y) \bigg[ \strucF{P_{\ell,m}\cos( - m\phi_h + m\phi_{R_\perp} + \phi_S)}{LT} \\
  &\quad \hspace*{0.4in} {} + \strucF{P_{\ell,m}\cos((2-m)\phi_h + m\phi_{R_\perp} - \phi_S)}{LT}\bigg]  \Bigg\} \; .
\end{split} 
\end{equation}
\end{widetext}

In the above equations, we can identify 21 different classes $F_{XY,Z}^{f(\phi_h, \phi_{R_\perp}, \phi_S)}$ 
of structure functions, for a total of 
\begin{equation}
21 \sum_{\ell=0}^{\ell_{\max}} (2\ell+1) = 21 (\ell_{\max}+1)^2
\end{equation}
structure functions. A common choice for dihadrons with invariant mass $M_h \lesssim 1$ GeV 
is to stop the sum at $\ell_{\max}=2$. The structure functions are then 189. 


\subsection{Integrated cross section}
\label{sec:IIC}

As already observed in Ref.~\cite{Diehl:2005pc}, the cross section for two-hadron
production integrated over the pair's transverse momentum has a similar form to the 
cross section of single-hadron production~\cite{Bacchetta:2006tn}, with $\phi_h$ replaced by 
$\phi_{R_\perp}$, i.e., 
\begin{widetext}
\begin{align}
\lefteqn{\frac{d\sigma}{d\xbj \, dy\, d\psi \,dz_h\, d\phi_{R_\perp}\, d M_{h}\,d \cos{\theta}} = 
\frac{\alpha^2}{\xbj y\, Q^2}\, \biggl( 1+\frac{\gamma^2}{2\xbj} \biggr) }  \nonumber \\ 
&  \quad \qquad \times \Biggl\{ 
A(x,y) \, F_{UU ,T} +  B(x,y) \, F_{UU ,L} + \frac{1}{2}\, V(x,y) \,\cos\phi_{R_\perp}\, F_{UU}^{\cos\phi_{R_\perp}} 
+ B(x,y) \, \cos(2\phi_{R_\perp})\,  F_{UU}^{\cos 2\phi_{R_\perp}} \nonumber \\  
& \qquad \qquad
+ \lambda_e\, \frac{1}{2}\, W(x,y) \,  \sin\phi_{R_\perp}\,  F_{LU}^{\sin\phi_{R_\perp}}
\phantom{\Bigg[ \Bigg] } \nonumber \\  
& \qquad \qquad
+ S_L\, \Bigg[  \frac{1}{2}\, V(x,y) \, \sin\phi_{R_\perp} \, F_{UL}^{\sin\phi_{R_\perp}}
+  B(x,y) \, \sin(2\phi_{R_\perp})\,  F_{UL}^{\sin 2\phi_{R_\perp}} \Bigg]  \nonumber \\  
& \qquad \qquad
+ S_L \lambda_e\, \Bigg[ \, C(x,y) \,  2 \, F_{LL}
+ \frac{1}{2}\, V(x,y)\, \cos\phi_{R_\perp}\, F_{LL}^{\cos \phi_{R_\perp}} \Bigg]  \nonumber \\  
& \qquad \qquad
+ |\bm{S}_\perp|\, \Bigg[ \sin(\phi_{R_\perp}-\phi_S)\, \Bigl( 
A(x,y) \, F_{UT ,T}^{\sin\left(\phi_{R_\perp} -\phi_S\right)}
+ B(x,y) \, F_{UT ,L}^{\sin\left(\phi_{R_\perp} -\phi_S\right)}\Bigr)  \nonumber \\  
& \qquad  \qquad \qquad
+ B(x,y) \, \sin(\phi_{R_\perp}+\phi_S)\,  F_{UT}^{\sin\left(\phi_{R_\perp} +\phi_S\right)}
+ B(x,y) \, \sin(3\phi_{R_\perp}-\phi_S)\,  F_{UT}^{\sin\left(3\phi_{R_\perp} -\phi_S\right)} 
\phantom{\Bigg[ \Bigg] } \nonumber \\  
& \qquad \qquad \qquad
+ \frac{1}{2}\, V(x,y) \,  \sin\phi_S\,  F_{UT}^{\sin \phi_S }
+ \frac{1}{2}\, V(x,y) \,  \sin(2\phi_{R_\perp}-\phi_S)\,  F_{UT}^{\sin\left(2\phi_{R_\perp} -\phi_S\right)}
\Bigg]  \nonumber \\  
& \qquad \qquad 
+ |\bm{S}_\perp| \lambda_e\, \Bigg[ C(x,y) \, \cos(\phi_{R_\perp}-\phi_S)\,  2\, F_{LT}^{\cos(\phi_{R_\perp} -\phi_S)}
+ \frac{1}{2}\, V(x,y) \, \cos\phi_S\, F_{LT}^{\cos \phi_S}  \nonumber \\  
& \qquad \qquad \qquad
+ \frac{1}{2}\, V(x,y)\,  \cos(2\phi_{R_\perp}-\phi_S)\,  F_{LT}^{\cos(2\phi_{R_\perp} - \phi_S)}
\Bigg] \Biggr\},
\label{eq:crossmasterint}
\end{align}
where the structure functions on the r.h.s.\ depend on $\xbj$, $Q^2$, $z_h$, $M_h$. 
The above formula can be obtained from the sum of the formulas in the previous section, 
observing that the only surviving contributions are the ones with values of $m$ that cancel 
the coefficients of the $\phi_h$ angle. Each of the 18 structure functions in Eq.~\eqref{eq:crossmasterint} 
corresponds to a specific class $F_{XY,Z}^{f(\phi_h, \phi_{R_\perp}, \phi_S)}$ in 
Eqs.~\eqref{eq:sigma_UU}-\eqref{eq:sigma_LT}. Only three classes do not survive the integration 
upon $P_{h \perp}$: the ones containing the fragmentation functions 
$\strucFpart{P_{\ell,m}\cos(m(\phi_h-\phi_{R_\perp}))}{LU,T}$ and 
$\strucFpart{P_{\ell,m}\cos(m(\phi_h-\phi_{R_\perp}))}{LU,L}$ in Eq.~\eqref{eq:sigma_LU}, and the one 
containing the fragmentation functions 
$\strucFpart{P_{\ell,m}\sin(-m\phi_h + m\phi_{R_\perp})}{UL}$ in Eq.~\eqref{eq:sigma_UL}.

\end{widetext}

\section{New definition for two-hadron fragmentation functions}
\label{sec:newDiFF}
\begin{figure}
\centering
\includegraphics[width=0.5\textwidth]{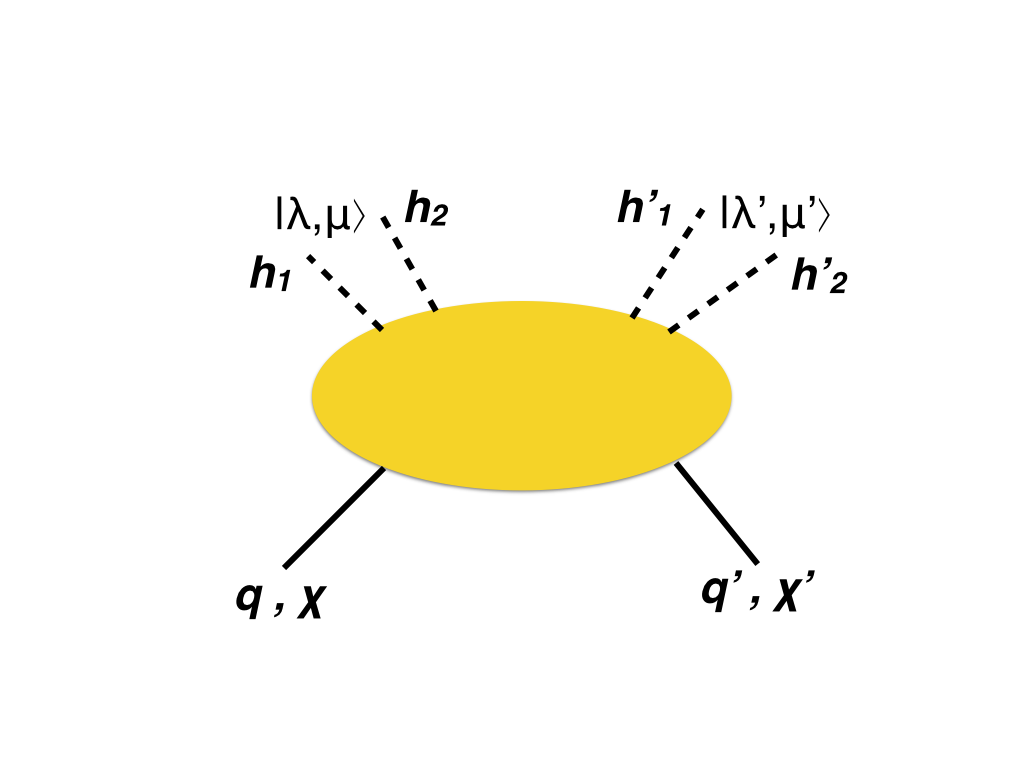}
\caption
[The generic diagram for leading-twist fragmentation functions.]  
{The generic diagram for leading-twist fragmentation functions.  The quarks are indicated as 
$q$, $q^\prime$, with their helicities $\chi$, $\chi^\prime$, respectively. The hadron pairs 
$(h_1, h_2)$ and $(h^\prime_1, h^\prime_2)$ are, respectively, in the partial waves $\ket{\lambda, \mu}$ and $\ket{\lambda^\prime,\mu^\prime}$.}
\label{fig:FF_diagram}
\end{figure}

Throughout the paper, we will adopt the following notation for intrinsic momenta: 
$k$ denotes the parton momentum in the distribution functions, 
$p$ the parton momentum occurring in the fragmentation functions, while 
$P$ refers to the final hadron momentum.

The general fragmentation process is rigorously defined starting from the correlation matrix 
$\Delta$~\cite{Mulders:1995dh}. Using the common shorthand notation for the trace of projections 
of the correlation matrix~\cite{Mulders:1995dh,Bianconi:1999cd}, 
\begin{equation}
\begin{array}{l}
  \Delta^{[\Gamma]}(z_h, \cos\vartheta, |\bm R_T|, |\bm p_T|, \bm R_T \cdot \bm p_T) =\hfill \\
 4 \pi \displaystyle{\frac{ z_h |\bm R|}{16 M_h}}  \int dk^+\,  \Tr{\Gamma\Delta(p,P_h,R)}\big|_{p^- =P_h^-/z} \; , 
\end{array}
\label{eq:DeltaGamma}
\end{equation}
with $p = \left\{ p^-, p^+ \equiv (p^2 + \bm p_T^2)/(2 p^-), \bm p_T\right\}$, for each choice 
of the operator $\Gamma$ we can define a specific class of fragmentation functions. 
At leading twist, we have the three classes 
\begin{eqnarray}
\label{eq:D1_def}
  D_1 &=& \Delta^{[\gamma^-]}\; , \\
\label{eq:G1_def}
  G_1 &=& \Delta^{[\gamma^- \gamma_5]}\; , \\
\label{eq:H1perp_def}
  i \frac{|\bm p_T|}{M_h} e^{i\phi_p} H_1^\perp &=& \Delta^{[-i(\sigma^{1-}+i\sigma^{2-})\gamma^5]}\nonumber\\
  &=& \Delta^{[(\gamma^2-i\gamma^1)\gamma^-\gamma^5]} \; ,
\end{eqnarray}
where $\phi_p$ is the azimuthal angle referred to $\bm{p}_T$~\cite{Bacchetta:2002fp}. 
Their probabilistic interpretation is depicted in Fig.~\ref{fig:FF_diagram}. With respect to the 
helicity density matrix of the fragmenting quark, the unpolarized $D_1$ corresponds to the sum 
of diagrams with no quark spin flip (those with $(\chi=\chi^\prime = 1/2)$ and $(\chi=\chi^\prime = -1/2)$ in 
Fig.~\ref{fig:FF_diagram}); the polarized $G_1$ corresponds to their difference; 
and the chiral-odd $H_1^\perp$, the generalized Collins fragmentation function, 
corresponds to the sum of diagrams with quark spin flip ($\chi \neq \chi^\prime$).  

One advantage of this new convention is that at leading twist it associates the name and symbol 
of the fragmentation functions with the quark spin states ($\chi$, $\chi^\prime$), while the various 
polarization states of the produced hadron system ($\ket{\lambda, \mu}$ and $\ket{\lambda^\prime, \mu^\prime}$) 
are associated with partial waves of the fragmentation functions. At subleading twist, the 
projection operators $\Gamma$ select the socalled "bad" light-cone components which are 
associated to quark-gluon combinations; hence, a clear identification of such components in terms 
of good quark helicity states is not applicable. 



\subsection{Partial wave expansion}
\label{sec:partial_wave_expansion}

The fragmentation functions can be expanded in partial waves in the direct product basis
$\ket{\lambda, \mu}\ket{\lambda^\prime, \mu^\prime}$, with $\lambda\, \lambda^\prime,$ the relative partial waves of each 
hadron pair in Fig.~\ref{fig:FF_diagram}. However, the structure functions in the cross section of 
Eqs.~\eqref{eq:sigma_UU}-\eqref{eq:sigma_LT} are related to specific partial waves in the direct 
sum basis. Then, it is more convenient to re-express the expansion in the basis $\ket{\ell, m}$, where 
$\ell = \lambda \oplus \lambda^\prime$ is the total partial wave of the hadronic system.

For each class of fragmentation functions in Eqs.~\eqref{eq:D1_def}-\eqref{eq:H1perp_def}, 
the partial wave expansion is accomplished by expanding the dependence on the polar angle 
$\cos\vartheta$ and on the azimuthal angle $\phi_{R_\perp} - \phi_p$ on a basis of spherical harmonics, 
in the same way as was done for the cross sections~\eqref{eq:sigma_UU}-\eqref{eq:sigma_LT}. 
Then,  we have
\begin{widetext}
   \begin{eqnarray}
      \label{eq:partial_wave_expansion_D1}
 D_1 &=& \sum_{\ell=0}^{\ell_{\max}}\sum_{m=-\ell}^\ell P_{\ell,m}( \cos\vartheta ) \, 
                     \cos \left( m \left(\phi_{R_\perp} -\phi_p \right)\right)  \, D^{\ket{\ell,m}}_1(z, M_{h}, |\bm p_T|), \\
      \label{eq:partial_wave_expansion_G1}
 G_1 &=& \sum_{\ell=1}^{\ell_{\max}}\sum_{m=-\ell}^\ell P_{\ell,m}( \cos\vartheta ) \, 
                     \sin \left( m \left(\phi_{R_\perp} -\phi_p \right)\right)  \, G^{\ket{\ell,m}}_1(z, M_{h}, |\bm p_T|), \\
      \label{eq:partial_wave_expansion_H1perp}
 H^{\perp}_1 &=& \sum_{\ell=0}^{\ell_{\max}}\sum_{m=-\ell}^\ell P_{\ell,m}(\cos\vartheta ) \, 
                     e^{im\left(\phi_{R_\perp} -\phi_p \right)} \, H^{\perp\ket{\ell,m}}_1(z, M_{h}, |\bm p_T|) \; , 
   \end{eqnarray}
\end{widetext}
and likewise for the higher twist fragmentation functions.  All non-expanded classes explicitly 
depend on the variables $z$, $M_h$, $|\bm p_T|$, $\cos\vartheta$, $\phi_{R_\perp}-\phi_p$,
and implicitly depend on $Q^2$.  



If the correlator $\Delta$ of Eq.~\eqref{eq:DeltaGamma} is considered as a Hermitean $2 \times 2$ matrix 
in the quark helicity basis, each diagonal element is complex conjugate of the other. Their sum gives twice 
their real part and is proportional to the class $D_1$; their difference gives twice their imaginary part and 
is proportional to the class $G_1$. Hence, when $D_1$ and $G_1$ are expanded onto the basis of 
spherical harmonics, the former contains only cosine components of the azimuthal angle 
$\phi_{R_\perp} - \phi_p$ while the latter only sine components. The class $H_1^{\perp}$, being 
related to helicity-flip matrix elements, contains both components. 

Each expanded fragmentation function appears in specific structure functions in 
Eqs.~\eqref{eq:sigma_UU}-\eqref{eq:sigma_LT} for a given partial wave 
$(\ell, m)$.


Note that the cross section for any final state polarization, when written in terms of 
non-expanded fragmentation functions, is identical to that for a single pseudo-scalar 
meson production, which is simply the $\ket{0,0}$ component of the final state polarization.  
This allows one to compute the cross section for any final state polarization, at
any twist level, given the cross section for pseudo-scalar meson production at the
corresponding twist level.

\section{Structure functions in terms of distribution and fragmentation functions}
\label{sec:struct_functs}
For a specific partial wave $(\ell, m)$, each structure function occurring in 
Eqs.~\eqref{eq:sigma_UU}-\eqref{eq:sigma_LT} can be written as a sum of 
convolutions with the form 
\begin{eqnarray}
\label{eq:mathcal_I}
  \mathcal{I}\left[ w f D \right] &=& \sum_q e_q^2 \int d^2 \bm k_T d^2 \bm p_T\, 
                        \delta^2\left( \bm k_T - \bm p_T - \frac{\bm P_{h \perp}}{z}\right) \nonumber \\
   & & \times \  w_m (x, z_h, M_h, \phi_h, \bm{k}_T, \bm{p}_T)  \nonumber \\ 
   & & \times \   x\, f^q(x, \bm{k}_T) \, 
           D^{q\, \ket{\ell, m}} (z_h, M_h, |\bm{p}_T|) \; , 
\end{eqnarray}
where the fragmentation functions $D^{q\, \ket{\ell, m}}$ have been defined in 
Eqs.~\eqref{eq:partial_wave_expansion_D1}-\eqref{eq:partial_wave_expansion_H1perp} 
for the leading twist case (and similarly for higher twists). 

\begin{widetext}
The leading twist structure functions for unpolarized beam and unpolarized target are
\begin{eqnarray}
  \label{eq:structfunc_UUL}
    \strucFpart{P_{\ell,m}\cos(m (\phi_h-\phi_{R_\perp}))}{UU,L} &=& 0 \; ,  \\
  \label{eq:structfunc_f1D1_UU}
    \strucFpart{P_{\ell,m}\cos(m(\phi_h-\phi_{R_\perp}))}{UU,T} &=& 
      \mathcal{I}\left[ \cos\big( m(\phi_h-\phi_p)\big) f_1 D_1^{\ket{\ell,m}} \right] \; ,  \\
  \label{eq:structfunc_cos2phi}
    \strucFpart{P_{\ell,m}\cos( (2-m)\phi_h+m\phi_{R_\perp})}{UU} &=& - \mathcal{I}\bigg[ 
      \frac{|\bm k_T| |\bm p_T |}{ M M_{h} }\, \cos\big( (m-2)\phi_h+\phi_k+(1-m)\phi_p\big)\, 
                   h_{1}^\perp H_1^{\perp \ket{\ell,m}} \bigg] \; , 
\end{eqnarray}
while the twist-3 structure functions are
\begin{eqnarray}
  \label{eq:structfunc_cosphi_UU}
    \strucFpart{P_{\ell,m}\cos((1-m)\phi_h+m\phi_{R_\perp})}{UU} &=& - \frac{2M}{Q} \mathcal{I}\Bigg[
      \frac{|\bm p_T|}{M_{h}}\cos\big( (m-1)\phi_h+(1-m)\phi_p\big) \nonumber \\ 
 && \hspace*{0.4in} {} \times \left(
      x h H_1^{\perp \ket{\ell,m}} + \frac{M_h}{M} f_1\frac{\tilde{D}^{\perp \ket{\ell,m}}}{z} \right) \nonumber \\
 && \hspace*{0.2in} {}  +\frac{|\bm k_T|}{M}\cos\big( (m-1)\phi_h+\phi_k-m\phi_p\big) \nonumber \\ 
 && \hspace*{0.4in} {} \times  \left(
      x f^\perp D_1^{\ket{\ell,m}} + \frac{M_h}{M} h_1^\perp \frac{\tilde{H}^{\ket{\ell,m}}}{z} \right)\Bigg] \; .
\end{eqnarray}

The leading twist structure functions for longitudinally polarized beam and unpolarized target are 
\begin{eqnarray}
  \label{eq:structfunc_f1D1_LUL}
    \strucFpart{P_{\ell,m}\sin(m(\phi_h-\phi_{R_\perp}))}{LU,L} &=& 0 \; ,  \\
  \label{eq:structfunc_f1D1_LU}
    \strucFpart{P_{\ell,m}\sin(m(\phi_h-\phi_{R_\perp}))}{LU,T} &=&-\mathcal{I}\left[ 
          2\cos\big( m (\phi_h-\phi_p) \big)  f_1 G_1^{\ket{\ell,m}} \right] \; ,
  \end{eqnarray}
while the twist-3 structure functions are 
\begin{eqnarray}
  \label{eq:structfunc_sinphi_LU}
    \strucFpart{P_{\ell,m}\sin((1-m)\phi_h+m\phi_{R_\perp})}{LU} &=& \frac{2M}{Q} \mathcal{I}\Bigg[
          -\frac{|\bm p_T|}{M_{h}}\cos\big( (1-m)\, (\phi_p- \phi_h) \big) \nonumber \\ 
&& \hspace*{0.5in} {} \times  \left(
        x e \, H_1^{\perp \ket{\ell,m}} + \frac{M_h}{M} f_1\frac{\tilde{G}^{\perp \ket{\ell,m}}}{z} \right) \nonumber \\
&& {}  +\frac{|\bm k_T|}{M}\cos\big( (m-1)\phi_h+\phi_k-m\phi_p\big) \nonumber \\ 
&& \hspace*{0.5in} {} \times \left(
      x g^\perp D_1^{\ket{\ell,m}} + \frac{M_h}{M} h_1^\perp \frac{\tilde{E}^{\ket{\ell,m}}}{z}  \right)\Bigg] \; .
\end{eqnarray}
 
The leading twist structure functions for unpolarized beam and longitudinally polarized target are
\begin{eqnarray}
  \label{eq:structfunc_sin2phi}
    \strucFpart{P_{\ell,m}\sin((2-m)\phi_h+m\phi_{R_\perp})}{UL} &=& -\mathcal{I}\bigg[ 
         \frac{|\bm k_T| |\bm p_T |}{ M M_{h} } \cos\big( (m-2)\phi_h+\phi_k+(1-m)\phi_p\big) \, 
               h_{1L}^\perp H_1^{\perp \ket{\ell,m}} \bigg] \; , \\
  \label{eq:structfunc_g1L_D1_UL}
    \strucFpart{P_{\ell,m}\sin(m(\phi_h-\phi_{R_\perp}))}{UL} &=&-\mathcal{I}\left[ 
        2\cos\big( m (\phi_h-\phi_p)\big) g_{1L} G_1^{\ket{\ell,m}} \right] \; , 
\end{eqnarray}
while the twist-3 structure functions are
\begin{eqnarray}
  \label{eq:structfunc_sinphi_UL}
    \strucFpart{P_{\ell,m}\sin((1-m)\phi_h+m\phi_{R_\perp})}{UL} &=& \frac{2M}{Q} \mathcal{I}\Bigg[
        -\frac{|\bm p_T|}{M_{h}}\cos\big( (1-m)\, (\phi_p-\phi_h)\big) \nonumber \\ 
&& \hspace*{0.5in} {} \times \left(
        x h_L H_1^{\perp \ket{\ell,m}} + \frac{M_h}{M} g_{1L}\frac{\tilde{G}^{\perp \ket{\ell,m}}}{z} \right) \nonumber \\
&& {}  +\frac{|\bm k_T|}{M}\cos\big( (m-1)\phi_h+\phi_k-m\phi_p\big) \nonumber \\ 
&& \hspace*{0.5in} {} \times \left(
        x f^\perp_L D_1^{\ket{\ell,m}} - \frac{M_h}{M} h_{1L}^\perp\frac{\tilde{H}^{\ket{\ell,m}}}{z} \right)\Bigg] \; .
  \end{eqnarray}
 
The leading twist structure functions for longitudinally polarized beam and longitudinally polarized target are  
\begin{eqnarray}
   \label{eq:structfunc_g1L_D1_LL}
      \strucFpart{P_{\ell,m}\cos(m(\phi_h-\phi_{R_\perp}))}{LL} &=& \mathcal{I}\left[ 
         \cos\big( m (\phi_h-\phi_p)\big)  g_{1L} D_1^{\ket{\ell,m}} \right] \; , 
  \end{eqnarray}
while the twist-3 structure functions are
\begin{eqnarray}
  \label{eq:structfunc_cosphi_LL}
    \strucFpart{P_{\ell,m}\cos((1-m)\phi_h+m\phi_{R_\perp})}{LL} &=& \frac{2M}{Q} \mathcal{I}\Bigg[
       \frac{|\bm p_T|}{M_{h}}\cos\big( (1-m)\, (\phi_p-\phi_h) \big) \nonumber \\ 
&& \hspace*{0.5in} {} \times \left(
       x e_L H_1^{\perp \ket{\ell,m}} - \frac{M_h}{M} g_{1L}\frac{\tilde{D}^{\perp \ket{\ell,m}}}{z} \right) \nonumber \\
&& {}  -\frac{|\bm k_T|}{M}\cos\big( (m-1)\phi_h+\phi_k-m\phi_p\big) \nonumber \\ 
&& \hspace*{0.5in} {} \times \left(
       x g^\perp_L D_1^{\ket{\ell,m}} + \frac{M_h}{M} h_{1L}^\perp\frac{\tilde{E}^{\ket{\ell,m}}}{z} \right)\Bigg] \; .
\end{eqnarray}

The leading twist structure functions for unpolarized beam and transversely polarized target are 
\begin{eqnarray}
    \strucFpart{P_{\ell,m}\sin((1+m)\phi_h-m\phi_{R_\perp}-\phi_S)}{UT,L} &=& 0 \; , \\ 
  \label{eq:structfunc_Sivers}
    \strucFpart{P_{\ell,m}\sin((1+m)\phi_h-m\phi_{R_\perp}-\phi_S)}{UT,T} &=& -\mathcal{I}\bigg[ 
      \frac{|\bm k_T|}{ M } \cos\big( \phi_k+m\phi_p-(1+m)\phi_h\big)  \nonumber \\ 
&& \hspace*{0.2in} {} \times 
       \left( f_{1T}^\perp D_1^{\ket{\ell,m}} + \textnormal{sign}[m] g_{1T} G_1^{\ket{\ell,m}} \right)\bigg] \; , \\
  \label{eq:structfunc_Collins}
    \strucFpart{P_{\ell,m}\sin((1-m)\phi_h+m\phi_{R_\perp}+\phi_S)}{UT} &=& -\mathcal{I}\bigg[ 
       \frac{|\bm p_T|}{ M_h } \cos\big( (1-m)\, (\phi_p-\phi_h) \big) \, h_1 H_1^{\perp \ket{\ell,m}}\bigg] \; , \\
  \label{eq:structfunc_pretzel}
    \strucFpart{P_{\ell,m}\sin((3-m)\phi_h+m\phi_{R_\perp}-\phi_S)}{UT} &=& - \mathcal{I}\bigg[ 
       \frac{|\bm k_T|^2 |\bm p_T |}{ 2 M^2 M_{h} } \cos\big( (m-3)\phi_h+2\phi_k+(1-m)\phi_p\big) \, h_{1T}^\perp 
            H_1^{\perp \ket{\ell,m}}\bigg] \; .
\end{eqnarray}
and at twist-3 the structure functions are
\begin{eqnarray}
  \label{eq:structfunc_UT_sinphi}
    \strucFpart{P_{\ell,m}\sin(\phi_S)}{UT} &=& \frac{2M}{Q} \mathcal{I}\Bigg\{
            \cos\big( m(\phi_h-\phi_p) \big)  \left( x f_T D_1^{\ket{\ell,m}} - \frac{M_{h}}{M}h_{1}\frac{\tilde{H}}{z}\right)
      \nonumber  \\ 
&&{} - \frac{1}{2}\frac{|\bm k_T| |\bm p_T|}{M M_h}\cos(m\phi_h+\phi_k-(m+1)\phi_p) \nonumber \\ 
&&\hspace*{0.5in}{} \times \Bigg[
          \left( x h_T H_1^{\perp \ket{\ell,m}} + \frac{M_h}{M}g_{1T} \frac{\tilde G^{\perp\ket{\ell,m}}}{z}\right)  \nonumber \\ 
&&\hspace*{1.0in}{}
       - \left( x h_T^\perp H_1^{\perp \ket{\ell,m}} - \frac{M_h}{M}f_{1T}^\perp\frac{\tilde D^{\perp\ket{\ell,m}}}{z}\right)\Bigg]\Bigg\} \end{eqnarray}
\begin{eqnarray}
  \label{eq:structfunc_UT_other_tw3}
     \strucFpart{P_{\ell,m}\sin((2-m)\phi_h+m\phi_{R_\perp}-\phi_S)}{UT} &=& \frac{2M}{Q}  \mathcal{I}\Bigg\{
           \frac{|\bm k_T|^2}{2M^2} \cos\big( (m-2)\phi_h+2\phi_k-m\phi_p\big) \nonumber \\ 
&& \hspace*{0.5in}{} \times
          \left( x f_T^\perp D_1^{\ket{\ell,m}} - \frac{M_{h}}{M}h^\perp_{1T}\frac{\tilde{H}}{z}\right)  \nonumber \\ 
&&{} - \frac{1}{2}\frac{|\bm k_T| |\bm p_T|}{M M_h}\cos\big( (m-2)\phi_h+\phi_k+(1-m)\phi_p\big) \nonumber \\ 
&&\hspace*{0.5in}{} \times \Bigg[
          \left( x h_T H_1^{\perp \ket{\ell,m}} + \frac{M_h}{M}g_{1T}\frac{\tilde G^{\perp\ket{\ell,m}}}{z}\right) \nonumber \\ 
&&\hspace*{1.0in}{}
          + \left( x h_T^\perp H_1^{\perp \ket{\ell,m}} - \frac{M_h}{M}f_{1T}^\perp\frac{\tilde D^{\perp\ket{\ell,m}}}{z}\right)\Bigg]
              \Bigg\} \; .
\end{eqnarray}

The leading twist structure functions for longitudinally polarized beam and transversely polarized target are 
\begin{eqnarray}
  \label{eq:structfunc_LT_tw2}
    \strucFpart{P_{\ell,m}\cos((1-m)\phi_h+m\phi_{R_\perp}-\phi_S)}{LT} &=& \mathcal{I}\bigg[ 
       \frac{|\bm k_T|}{ M } \cos\big( (m-1)\phi_h+\phi_k-m\phi_p\big)   \nonumber \\ 
&& \hspace*{0.2in} {} \times 
       \left( g_{1T} D_1^{\ket{\ell,m}} + \textnormal{sign}[m] f_{1T}^\perp G_1^{\ket{\ell,m}}\right)\bigg] \; , 
\end{eqnarray}
while the twist-3 structure functions are
\begin{eqnarray}
  \label{eq:structfunc_LT_cosphi}
    \strucFpart{P_{\ell,m}\cos(\phi_S)}{LT} &=& \frac{2M}{Q} \mathcal{I}\Bigg\{
          -\cos\big( m(\phi_h-\phi_p)\big) \,  \left( x g_T D_1^{\ket{\ell,m}} + \frac{M_{h}}{M}h_{1}\frac{\tilde{E}}{z}\right) \nonumber \\
&&{} + \frac{1}{2}\frac{|\bm k_T| |\bm p_T|}{M M_h}\cos\big( m\phi_h+\phi_k-(m+1)\phi_p\big) \nonumber \\ 
&&\hspace*{0.5in}{} \times \Bigg[
         \left(x e_T H_1^{\perp \ket{\ell,m}} - \frac{M_h}{M}g_{1T}\frac{\tilde D^{\perp\ket{\ell,m}}}{z}\right) \nonumber \\ 
&&\hspace*{1.0in}{}
       + \left( x e_T^\perp H_1^{\perp \ket{\ell,m}} + \frac{M_h}{M}f_{1T}^\perp\frac{\tilde G^{\perp\ket{\ell,m}}}{z}\right)\Bigg]
          \Bigg\} \; ,
\end{eqnarray}
\begin{eqnarray}
  \label{eq:structfunc_LT_other_tw3}
    \strucFpart{P_{\ell,m}\cos((2-m)\phi_h+m\phi_{R_\perp}-\phi_S)}{LT} &=& \frac{2M}{Q} \mathcal{I}\Bigg\{
           -\frac{|\bm k_T|^2}{2M^2} \cos\big( (m-2)\phi_h+2\phi_k-m\phi_p\big) \nonumber \\ 
&&\hspace*{0.5in}{} \times
          \left( x g_T^\perp D_1^{\ket{\ell,m}} + \frac{M_{h}}{M}h^\perp_{1T}\frac{\tilde{E}}{z}\right) \nonumber \\ 
&&{} + \frac{1}{2}\frac{|\bm k_T| |\bm p_T|}{M M_h}\cos\big( (m-2)\phi_h+\phi_k+(1-m)\phi_p\big) \nonumber \\ 
&&\hspace*{0.5in}{} \times \Bigg[
         \left( x e_T H_1^{\perp \ket{\ell,m}} - \frac{M_h}{M}g_{1T}\frac{\tilde D^{\perp\ket{\ell,m}}}{z}\right) \nonumber \\
&&\hspace*{1.0in}{}
       - \left( x e_T^\perp H_1^{\perp \ket{\ell,m}} + \frac{M_h}{M}f_{1T}^\perp\frac{\tilde G^{\perp\ket{\ell,m}}}{z}\right)\Bigg]
           \Bigg\} \; .
\end{eqnarray}

\end{widetext}

\section{Relation with existing nomenclature}
\label{sec:specific_final_states}
If the final hadronic system is made of pairs of mesons, then in Fig.~\ref{fig:FF_diagram} 
we can have $\lambda, \, \lambda^\prime = 0,1$. Then, in the direct sum basis 16 states 
can be formed:
\begin{equation}
\label{eq:2mesons_usual_coupling}
  \left( 1 \oplus 0 \right) \otimes \left( 1 \oplus 0 \right) = 2 \oplus 1 \oplus 1 \oplus 1 \oplus 0 \oplus 0\; . 
\end{equation}
In reality, the cross section appears 
as if there were 9 states with $\ell$ taking the values $\ell = 0, 1, 2$, where the $\ell = 1$ and 
$\ell = 0$ states are three times and two times degenerate, respectively. The $\ell = 0$ states 
can be distinguished, because one of them is the cross section for a single pseudo-scalar meson 
production, while the other one is the angular integrated dihadron cross section. The $\ell=1$ states 
are not experimentally distinguishable. They contain a contribution from the interference of 
relative partial waves $s$ and $p$ ($[\lambda = 0] \otimes [\lambda^\prime = 1]$ and viceversa), 
and from the interference of two $p$ waves ($[\lambda = 1] \otimes [\lambda^\prime = 1]$), 
in agreement with Ref.~\cite{Bacchetta:2002ux}. 

We now clarify our notation by recovering known results in the literature for specific final hadronic 
systems. 


\subsection{Single-hadron SIDIS}
\label{sec:1hSIDIS}

For the production of a pseudo-scalar meson, only the $\ket{0,0}$ final state polarization is possible.  
Hence, for the case $\ell = 0$, $m=0$, the cross sections~\eqref{eq:sigma_UU}-\eqref{eq:sigma_LT} 
with the structure functions~\eqref{eq:structfunc_UUL}-\eqref{eq:structfunc_LT_other_tw3} 
reduce to the ones in Ref.~\cite{Bacchetta:2006tn} with the following obvious identifications:
\begin{align}
D_1^{\ket{0,0}} &= D_1 \; , &H_1^{\perp \ket{0,0}} &= H_1^\perp \; , &\tilde{H}^{\ket{0,0}} &= \tilde{H} \; , \\
\tilde{D}^{\perp \ket{0,0}} &= \tilde{D}^\perp \; , &\tilde{G}^{\perp \ket{0,0}} &= \tilde{G}^\perp \; , &\tilde{E}^{\ket{0,0}} &= \tilde{E} \; . 
\label{eq:1hcorresp}
\end{align}
The $D_1$ and $H_1^{\perp}$ are the usual unpolarized fragmentation function and the Collins 
function, respectively, and correspond to the reduction of 
Eqs.~\eqref{eq:partial_wave_expansion_D1}-\eqref{eq:partial_wave_expansion_H1perp} to the case 
$\ell = 0$, $m=0$. In this limit, no contribution emerges from the class of fragmentation functions $G_1$.


\subsection{Two-hadron SIDIS}
\label{sec:2hSIDIS}

If the final state is represented by two mesons, the crosscheck with existing literature can be made 
by comparing Eqs.~\eqref{eq:sigma_UU}-\eqref{eq:sigma_LT}, including the leading twist 
contributions to the structure functions of Eqs.~\eqref{eq:structfunc_UUL}-\eqref{eq:structfunc_LT_other_tw3}, 
with Eqs.~(C4)-(C10) of Ref.~\cite{Bacchetta:2002ux}. 
For the chiral-even fragmentation functions we have
\begin{align}
   D^{\ket{0,0}}_1 &=  \frac{1}{4} D_{1,OO}^{s} + \frac{3}{4} D_{1,OO}^{p} \; , 
\end{align}
\begin{align}
   D^{\ket{1,0}}_1 &= D_{1,OL} \; , &D^{\ket{1,1}}_1 = D^{\ket{1,-1}}_1 = \frac{1}{2}\, D_{1,OT} \; , \\
   D^{\ket{2,0}}_1 &= \frac{1}{2}\, D_{1,LL} \; , &D^{\ket{2,1}}_1 = D^{\ket{2,-1}}_1 = \frac{1}{4}\, D_{1,LT} \; , \\
   & &D^{\ket{2,2}}_1 = D^{\ket{2,-2}}_1 = \frac{1}{2}\, D_{1,TT} \; , 
\end{align}
\begin{align}
   G^{\ket{0,0}}_1 &= G^{\ket{1,0}}_1 = G^{\ket{2,0}}_1 = 0 \; , \\
   G^{\ket{1,1}}_1 &= G^{\ket{1,-1}}_1 = - \frac{\left|\bm p_T \right| |\bm R|}{2 M_{h}^2} G_{1,OT}^{\perp} \; , \\
   G^{\ket{2,1}}_1 &= G^{\ket{2,-1}}_1 = - \frac{\left|\bm p_T \right| |\bm R|}{4 M_{h}^2} G_{1,LT}^{\perp} \; , \\
   G^{\ket{2,2}}_1 &= G^{\ket{2,-2}}_1= - \frac{\left|\bm p_T \right| |\bm R|}{4 M_{h}^2} G_{1,TT}^{\perp} \; ,
\label{eq:2hchevencorresp}
\end{align}
while for the chiral-odd function,
\begin{align}
   H^{\perp \ket{0,0}}_1 &= \frac{1}{4} H_{1,OO}^{\perp s} + \frac{3}{4} H_{1,OO}^{\perp p} \; , \\
   H^{\perp \ket{1,0}}_1 &= H_{1,OL}^\perp \; , \quad H^{\perp \ket{2,0}}_1 = \frac{1}{2} H_{1,LL}^{\perp} \; ,
\end{align}
\begin{align}
   H^{\perp \ket{1,1}}_1 &= \frac{|\bm R|}{\left| \bm p_T \right|} H_{1,OT}^{\varangle}  \; ,  
      &H^{\perp \ket{1,-1}}_1 = H_{1,OT}^{\perp} \; , \\
   H^{\perp \ket{2,1}}_1 &= \frac{|\bm R|}{2 \left| \bm p_T \right|} H_{1,LT}^{\varangle} \; , 
       &H^{\perp \ket{2,-1}}_1 = \frac{1}{2} H_{1,LT}^{\perp} \; , \\
   H^{\perp \ket{2,2}}_1 &= \frac{|\bm R|}{\left| \bm p_T \right|} H_{1,TT}^{\varangle} \; , 
       &H^{\perp \ket{2,-2}}_1 = H_{1,TT}^{\perp} \; . 
\label{eq:2hchevencorresp}
\end{align}

Using the above relations, one can then cross-check the formulae listed in 
Secs.~\ref{sec:cross_section} and~\ref{sec:struct_functs}. There is consistency 
between the published literature and the present work, although in some case 
there are typographical errors (for a detailed list, see Appendix~\ref{sec:appendixB}). 

\section{Conclusion}
\label{sec:conclusion}
In this paper, we have presented a slightly modified definition of the fragmentation functions 
compared to, e.g., Ref.~\cite{Bacchetta:2003vn}. We have proposed a new partial wave expansion 
for fragmentation functions, which allows a consistent framework for fragmentation into
final states of any polarization. 

This not only helps in the interpretation of cross section moments, but also has the advantage 
that the two-hadron SIDIS cross sections, at any twist, can be derived 
from single-hadron SIDIS. Using this method, in this paper we present for the first time the expression 
of the two-hadron SIDIS cross section up to subleading twist, including the dependence upon the 
transverse momentum of involved particles. 

The cross section has also been given in terms of structure functions, and the resulting expressions 
have been cross-checked with existing literature for specific cases.

\section{Acknowledgments}
We thank Markus Diehl for helpful discussions. S.~G. gratefully acknowledges the mentoring 
of Wolfgang Lorenzon and Hal Spinka.

This work was partially supported: by the National Science Foundation, grant number 0855368; 
by the Nuclear Physics Program of the US Department of Energy, Office of Science to the
Medium Energy Nuclear Physics group, High Energy Physics Division, Argonne National Laboratory; 
by the European Community through the Research Infrastructure Integrating Activity ``HadronPhysics3'' 
(Grant Agreement n. 283286) under the European 7th Framework Programme

\appendix
\section{Definition of azimuthal angles}
\label{sec:appendixA}
As explained in Sec.~\ref{sec:IIA}, the SIDIS cross section for dihadron production depends also 
on the azimuthal angle $\phi_{R_\perp}$ of the vector $R_T$ measured in the plane 
perpendicular to $(P,q)$, where $R_T$ is given by Eq.~\eqref{eq:RT} and $\phi_{R_\perp}$ 
is defined by 
\begin{align}
  \label{eq:phi-Rperp}
\cos\phi_{R_\perp} &= - \frac{l_\mu R_{T \nu}\, g_\perp^{\mu\nu}}{%
  \sqrt{ l_\perp^2\, R_{T \perp}^2}} \,,
&
\sin\phi_{R_\perp} &= - \frac{l_\mu R_{T \nu}\, \epsilon_\perp^{\mu\nu}}{%
  \sqrt{ l_\perp^2\, R_{T \perp}^2}} \, ,
\end{align}
with $l_\perp^\mu = g_\perp^{\mu\nu} l^{}_\nu$ and $R_{T \perp}^\mu = g_\perp^{\mu\nu} R^{}_{T \nu}$. 

Depending on the reference frame, the vector $R_T$ can have a non-vanishing component 
along $\bm q$, but $g_\perp^{\mu\nu}$ projects out only its spatial components transverse to $\bm q$. 
Hence, in order to compare with other non-covariant definitions we inspect in the following 
the expressions of only $\bm{R}^{}_{T \perp} \equiv \{ R^{}_{T x}, \, R^{}_{T y} \}$. 

The most natural choice of frame is the Target Rest Frame (TRF). There, from Eq.~\eqref{eq:RT} we have
\begin{equation}
\bm{R}^{}_{T \perp} \Big\vert_{\mathrm{TRF}} = \frac{z_2 \bm{P}_{1 T} - z_1 \bm{P}_{2 T}}{z} + {\cal O} \left(
 \frac{1}{Q^3} \right) \, .
\label{eq:RT-TRF}
\end{equation}
The above result coincides (up to corrections of order $1/Q^3$) with the transverse spatial components 
of $\bm{R} - \bm{P}_h \  \bm{R}\cdot \bm{P}_h / \bm{P}_h^2$, which is the definition of $\bm{R}_T$ 
used in the analysis of dihadron production from SIDIS data by the HERMES 
Collaboration~\cite{Airapetian:2008sk}. It is also equal, in the same limit, to the definition used in 
Ref.~\cite{Artru:1995zu}, that has been adopted in the analyses of dihadron production from SIDIS 
data by the COMPASS Collaboration~\cite{Adolph:2012nw} and from $e^+ e^-$ annihilation data 
by the Belle Collaboration~\cite{Vossen:2011fk}. 

If we boost all four-vectors to the so-called Infinite Momentum Frame (IMF), where the momentum of the 
virtual photon is purely space-like, our definition reduces to 
\begin{equation}
\bm{R}^{}_{T \perp} \Big\vert_{\mathrm{IMF}} = \frac{z_2 \bm{P}_{1 T} - z_1 \bm{P}_{2 T}}{z} + {\cal O} \left(
 \frac{1}{Q^2} \right) \, ,
\label{eq:RT-IMF}
\end{equation}
which again coincides with all other non-covariant definitions, but now up to corrections of order $1/Q^2$. 
We find the same result if we boost the four-vectors to the Breit frame of the virtual photon-proton system, 
{\it i.e.} where the vector $q+P$ is purely time-like. 

In conclusion, we find that our definition of the azimuthal angle $\phi_{R_\perp}$ of the vector 
$R_{T \perp}^\mu = g_\perp^{\mu\nu} R^{}_{T \nu}$, with $R_T$ given in Eq.~\eqref{eq:RT}, is equivalent 
to all other definitions found in the literature up to corrections of order $1/Q^2$. In the target rest frame, 
the equivalence with the definition of Ref.~\cite{Airapetian:2008sk} holds including $1/Q^2$ corrections, 
{\it i.e.} up to correction of order $1/Q^3$. Of course, our definition is preferable because it is covariant, 
hence valid in any frame. Recently, a new definition appeared~\cite{Kotzinian:2014lsa} which in the 
notation of this paper reads $\bm{R}_\perp = (\bm{P}_{1 \perp} - \bm{P}_{2 \perp})/2$; this 
definition is different from all the other ones.

\section{Crosscheck of structure functions}
\label{sec:appendixB}
As explained in Sec.~\ref{sec:2hSIDIS}, the formulae for the cross sections and structure functions 
listed in Secs.~\ref{sec:cross_section} and \ref{sec:struct_functs}, respectively, can be cross-checked 
at the leading twist level with Eqs.~(C4)-(C10) of Ref.~\cite{Bacchetta:2002ux}. In the crosscheck, 
the different convention in the definition of azimuthal angles must be accounted for, because 
Ref.~\cite{Bacchetta:2002ux} was published before the release of the so-called Trento 
conventions~\cite{Bacchetta:2004jz}. There is a general consistency between the two groups of 
equations, but the latter one displays some typographical errors that are listed here below.

For unpolarized beam and target, the cross section $d\sigma_{UU}$ of Eq.~(C4) can be divided in 
two groups enclosed in braces, the former multiplied by $A(y)$ and the latter by $B(y)$. 
The third term of the former group, involving the function $D_{1,OT}$ and corresponding to the 
component $\ell = 1$, $m= \pm 1$ in Eq.~\eqref{eq:sigma_UU}, should change sign. 
In the latter group, a term proportional to
\begin{equation}
\cos\vartheta \cos 2\phi_h \mathcal{I} \left[ 
\frac{2 \bm{k}_T\cdot \hat{\bm{P}}_{h\perp}\, \bm{p}_T\cdot \hat{\bm{P}}_{h\perp} - \bm{k}_T \cdot \bm{p}_T}
        {M M_h} \, h_1^\perp \, H_{1,OL}^\perp \right]  \nonumber
\end{equation}
is missing, that corresponds to the component $\ell = 1$, $m= 0$ in Eq.~\eqref{eq:sigma_UU}. 

For unpolarized beam and longitudinally polarized target, the cross section $d\sigma_{UL}$ of Eq.~(C6) 
can also be divided in two groups enclosed in braces, the former multiplied by $A(y)$ and the latter by $B(y)$. 
The overall $(-)$ sign in front of the former group should be dropped. The first term of the latter group, 
involving $\frac{1}{4} H_{1,OO}^{\perp s} + \frac{3}{4} H_{1,OO}^{\perp p}$ and corresponding to the 
component $\ell = 0$, $m= 0$ in Eq.~\eqref{eq:sigma_UL}, should change sign. Finally, a term 
proportional to 
\begin{equation}
\cos\vartheta \sin 2\phi_h \mathcal{I} \left[ 
\frac{2 \bm{k}_T\cdot \hat{\bm{P}}_{h\perp}\, \bm{p}_T\cdot \hat{\bm{P}}_{h\perp} - \bm{k}_T \cdot \bm{p}_T}
        {M M_h} \, h_{1L}^\perp \, H_{1,OL}^\perp \right]  \nonumber
\end{equation}
is missing, that corresponds to the component $\ell = 1$, $m= 0$ in Eq.~\eqref{eq:sigma_UL}. 

For longitudinally polarized beam and target, in the cross section $d\sigma_{LL}$ of Eq.~(C7) 
the fourth term proportional to $\sin\vartheta \cos (\phi_h - \phi_{R_\perp})$ should involve the 
function $D_{1,OT}$. It corresponds to the component $\ell = 1$, $m= 1$ in Eq.~\eqref{eq:sigma_LL}.

For unpolarized beam and transversely polarized target, the cross section $d\sigma_{UT}$ of Eq.~(C8)
can also be divided in two groups enclosed in braces, the former multiplied by $A(y)$ and the latter by $B(y)$. 
The eighth term of the former group, involving the function $D_{1,OL}$ and corresponding to the 
component $\ell = 1$, $m= 0$ in Eq.~\eqref{eq:sigma_UT}, should read
\begin{equation}
\cos\vartheta \sin (\phi_h - \phi_S) \mathcal{I} \left[ 
\frac{\bm{k}_T\cdot \hat{\bm{P}}_{h\perp}}{M} \, f_{1T}^\perp \, D_{1,OL} \right] \; . \nonumber
\end{equation}
In the latter group, the first and ninth terms, corresponding to the component $\ell = 0$, $m= 0$ 
in Eq.~\eqref{eq:sigma_UT}, should read
\begin{widetext}
\begin{eqnarray}
& &\sin (\phi_h + \phi_S) \mathcal{I} \left[  \frac{\bm{p}_T\cdot \hat{\bm{P}}_{h\perp}}{M_h} \, 
h_1 \, \left( \frac{1}{4} H_{1,OO}^{\perp s} + \frac{3}{4} H_{1,OO}^{\perp p} \right) \right] \; , \nonumber \\
& &- \sin (3 \phi_h - \phi_S) \mathcal{I} \left[  
\frac{4 (\bm{k}_T\cdot \hat{\bm{P}}_{h\perp})^2\, \bm{p}_T\cdot \hat{\bm{P}}_{h\perp} - 
          2 \bm{k}_T\cdot \hat{\bm{P}}_{h\perp} \, \bm{k}_T\cdot \bm{p}_T - 
           \bm{k}_T^2 \, \bm{p}_T\cdot \hat{\bm{P}}_{h\perp}}{M^2 M_h} \, h_{1T}^\perp \, 
           \left( \frac{1}{4} H_{1,OO}^{\perp s} + \frac{3}{4} H_{1,OO}^{\perp p} \right) \right] \; . 
 \nonumber
\end{eqnarray}
Finally, the terms
\begin{eqnarray}
& &- \cos\vartheta \sin (\phi_h + \phi_S) \mathcal{I} \left[  \frac{\bm{p}_T\cdot \hat{\bm{P}}_{h\perp}}{M_h} \, 
h_1 \, H_{1,OL}^{\perp} \right] \; , \nonumber \\
& &\cos\vartheta \sin (3 \phi_h - \phi_S) \mathcal{I} \left[  
\frac{4 (\bm{k}_T\cdot \hat{\bm{P}}_{h\perp})^2\, \bm{p}_T\cdot \hat{\bm{P}}_{h\perp} - 
          2 \bm{k}_T\cdot \hat{\bm{P}}_{h\perp} \, \bm{k}_T\cdot \bm{p}_T - 
           \bm{k}_T^2 \, \bm{p}_T\cdot \hat{\bm{P}}_{h\perp}}{M^2 M_h} \, h_{1T}^\perp \, 
           H_{1,OL}^{\perp} \right] \; , \nonumber
\end{eqnarray}
\end{widetext}
are missing, that correspond to the component $\ell = 1$, $m= 0$ in Eq.~\eqref{eq:sigma_UT}. 

For longitudinally polarized beam and transversely polarized target, in the cross section 
$d\sigma_{LT}$ of Eq.~(C9) the terms 
\begin{widetext}
\begin{eqnarray}
& &- \cos\vartheta \cos (\phi_h - \phi_S) \mathcal{I} \left[ \frac{\bm{k}_T\cdot \hat{\bm{P}}_{h\perp}}{M} \, 
g_{1T} \, D_{1,OL} \right] \; , \nonumber \\
& &- \sin\vartheta \cos (\phi_{R_\perp} - \phi_S) \mathcal{I} \left[ \frac{\bm{k}_T\cdot \bm{p}_T}{M} \, 
\left( g_{1T} \, \frac{1}{2 |\bm{p}_T|} D_{1,OT} - f_{1T}^\perp \, \frac{|\bm{R}|}{2 M_h^2} G_{1,OT}^\perp \right) \right] 
\; , \nonumber \\
& &- \sin\vartheta \cos (2 \phi_h -\phi_{R_\perp} - \phi_S) \mathcal{I} \left[ 
\frac{2 \bm{k}_T\cdot \hat{\bm{P}}_{h\perp}\, \bm{p}_T\cdot \hat{\bm{P}}_{h\perp} - \bm{k}_T \cdot \bm{p}_T}
        {M}  \, \left( g_{1T} \, \frac{1}{2 |\bm{p}_T|} D_{1,OT} + f_{1T}^\perp \, \frac{|\bm{R}|}{2 M_h^2} G_{1,OT}^\perp \right) \right] 
\; , \nonumber 
\end{eqnarray}
\end{widetext}
are missing, that correspond to the components $\ell = 1$, $m= 0$, and $\ell = 1$, $m= 1$, and 
$\ell = 1$, $m= -1$, respectively, in Eq.~\eqref{eq:sigma_LT}.

\bibliography{mybiblio}

\end{document}